\documentclass[fleqn,twoside]{article}
\usepackage{espcrc2}

\usepackage{graphicx}
\usepackage[figuresright]{rotating}

\newcommand{\AmS}{{\protect\the\textfont2
  A\kern-.1667em\lower.5ex\hbox{M}\kern-.125emS}}

\title{Physics Potential of Solar Neutrino Experiments}

\author{A.B. Balantekin\address[MCSD]{University of Wisconsin, Department of Physics,  
	Madison, WI  53706  USA}%    
        \thanks{\tt baha@nucth.physics.wisc.edu},
        H. Y\"uksel\addressmark\thanks{\tt yuksel@nucth.physics.wisc.edu}}

\begin{document}

\begin{abstract}
We discuss the physics potential of the solar neutrino experiments i) To explore the 
parameter space of neutrino mass and mixings; ii) To probe the physics of the Sun; 
iii) To explore nuclear physics of the neutrino-target interactions. Examples are given 
for these three classes. 
\vspace{1pc}
\end{abstract}

\maketitle

\begin{figure}
\includegraphics[scale=0.35]{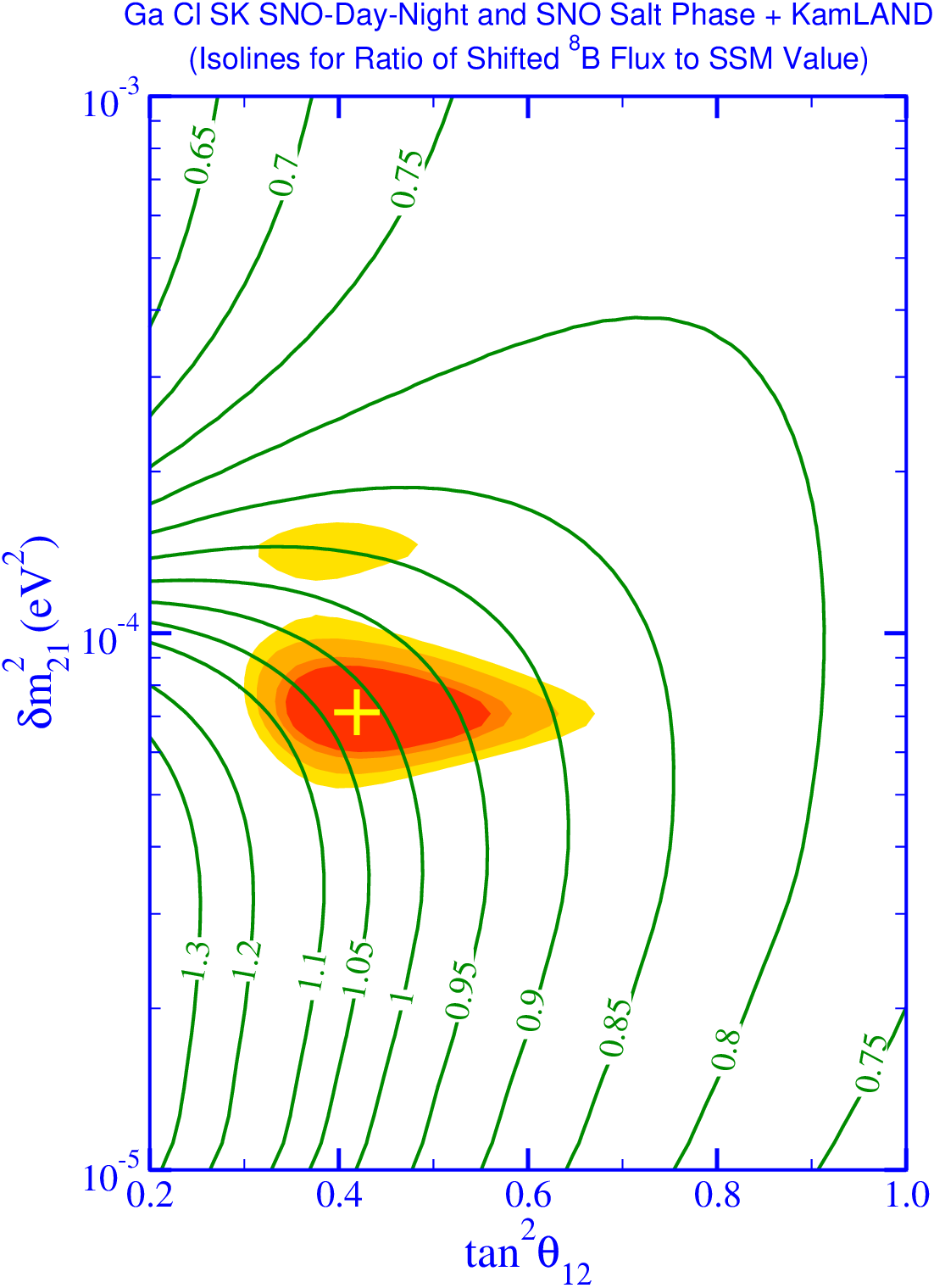}
\vspace*{0cm} \caption{ \label{fig:1}
Allowed confidence levels from the joint analysis of all
available solar neutrino data (chlorine, average gallium, SNO and SK
spectra and SNO salt phase) and KamLAND reactor data
The isolines are the ratio of the shifted $^8$B flux 
to the SSM value. 
At best fit (marked by a cross) the value of this ratio is $1.02$ 
(from Ref. \cite{Balantekin:2003jm}). 
}\end{figure}

\begin{figure}[t]
\includegraphics[scale=0.29]{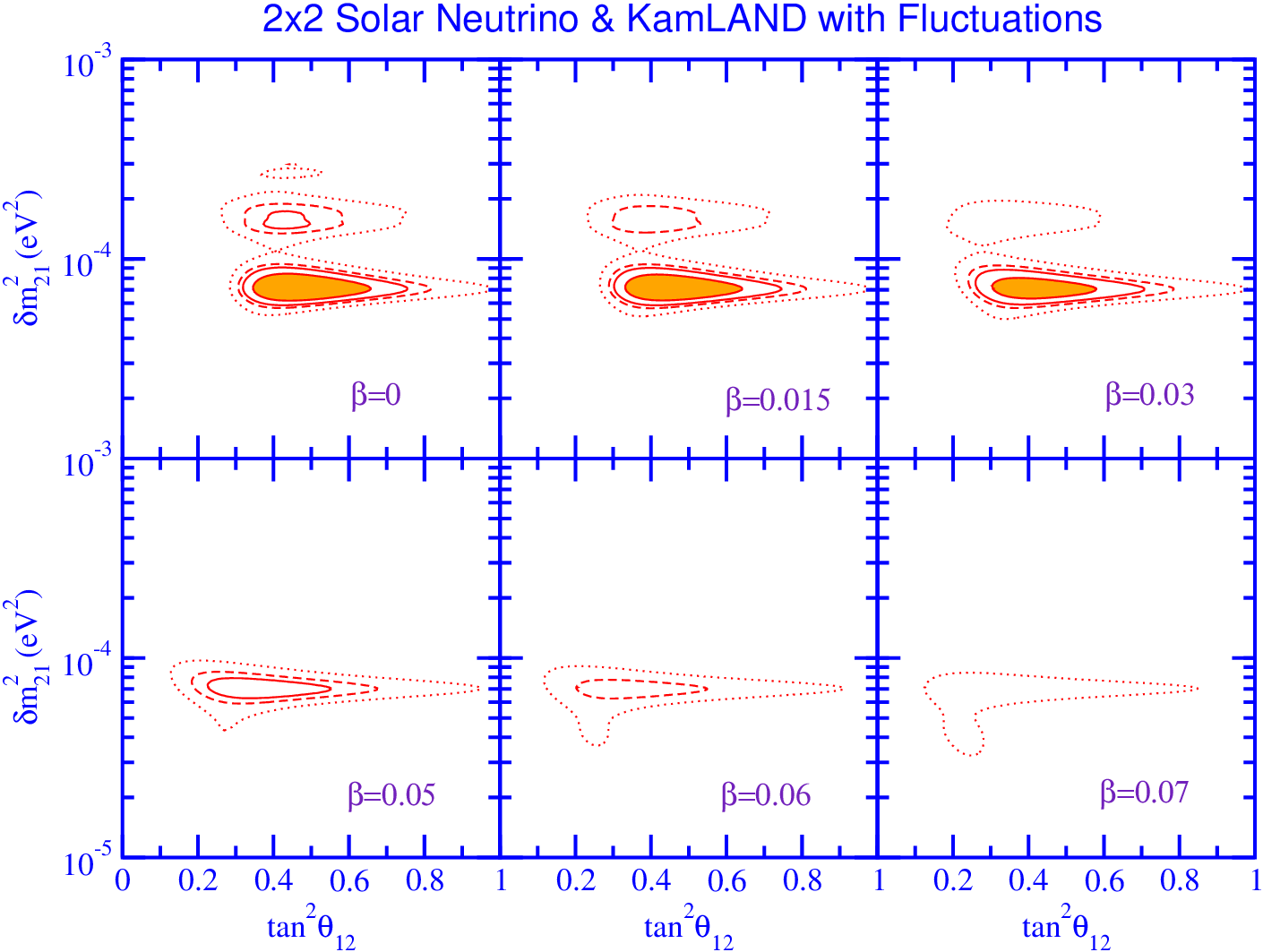}
\vspace*{0cm} \caption{ \label{fig:2}  Allowed regions of the
  neutrino parameter space with solar-density fluctuations when the
  data from the solar neutrino and KamLAND experiments are used. 
  The SSM density profile of
  Ref. \cite{Bahcall:2000nu} and the correlation length of 10 km are
  used. The case with no fluctuations ($\beta=0$) are compared with
  results obtained with the indicated fractional fluctuation.  The
  shaded area is the 70 \% confidence level region. 90 \% (solid
  line), 95 \% (dashed line), and 99 \% (dotted line) confidence
  levels are also shown (From Ref. \cite{Balantekin:2003qm}).}
\end{figure}

\begin{figure}[t]
\includegraphics[scale=0.3]{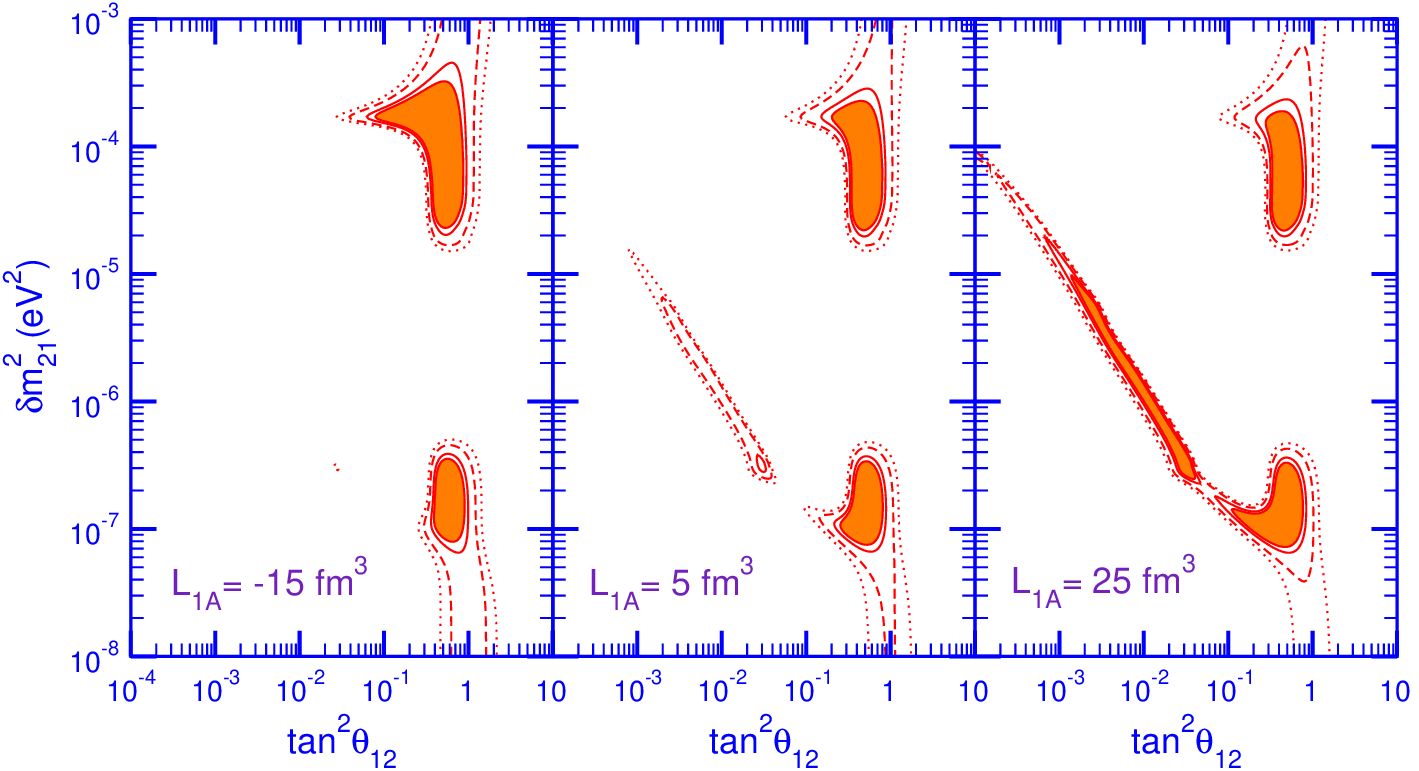}
\vspace*{0cm} \caption{ \label{fig:3} 
The change in the allowed region of the neutrino parameter space
using solar neutrino data measured at SNO as the value of
$ L_{1A} $ changes. The shaded
areas are the 90 \% confidence level region. 95 \% (solid line),
99 \% (log-dashed line), and 99.73 \% (dotted-line) confidence
levels are also shown (From Ref. \cite{Balantekin:2003ep}).}
\end{figure}

Recently announced results from the Sudbury Neutrino Observatory (SNO) 
\cite{Ahmed:2003kj} and KamLAND \cite{Eguchi:2002dm} experiments 
indicate that 
neutrino physics is moving from the discovery stage to the precision measurements 
stage. (For recent reviews see e. g. \cite{Smirnov:2003xe,Barger:2003qi}). A combined 
analysis of the data from these experiments as well as data from other solar neutrino 
experiments (Super-Kamiokande [SK] 
\cite{Fukuda:2002pe}, Chlorine \cite{Cleveland:nv}, and 
Gallium \cite{Abdurashitov:2002nt,Hampel:1998xg,Altmann:2000ft}), place severe 
constraints on the neutrino parameters, especially mixing between first and second 
generations \cite{Balantekin:2003dc,deHolanda:2003nj,Balantekin:2003jm}. 
As an example the neutrino parameter space obtained from the global analysis of all 
available solar neutrino plus the KamLAND
data is shown in Fig. \ref{fig:1} \cite{Balantekin:2003jm}. 

The aim of this short contribution is to re-emphasize that, in principle, 
high-precision solar-neutrino data have potential beyond exploring neutrino 
parameter space. Here we consider two other applications to solar physics and 
to nuclear physics of the neutrino-target interactions.

Some time ago it was pointed out that solar neutrino data can be inverted 
to extract information about the density scale height \cite{Balantekin:1997fr} 
in a similar way the 
helioseismological
information is inverted to obtain  the sound-speed profile
throughout the Sun. Even though the precision of the data has not yet reached to a 
point where such an inversion is possible, it is currently possible to obtain rather 
tight limits on {\em fluctuations} of the solar density. 
To do so one assumes \cite{Loreti:1994ry} that 
the electron density $N_e$ fluctuates around the
value, $\langle N_e \rangle$, predicted by the Standard Solar Model (SSM) 
\cite{Bahcall:2000nu}
\begin{equation}
N_e (r) = (1 + \beta F (r)) \langle N_e (r) \rangle ,
\label{flucdef}
\end{equation}
and that the fluctuation $F (r)$ takes the form of 
white-noise. The neutrino parameter space 
for various values of the parameter $\beta$ was calculated in Ref. 
\cite{Balantekin:2003qm} and is shown in Figure 2. These results, in agreement with 
the calculations of other authors \cite{Burgess:2003su,Guzzo:2003xk}, show that 
the neutrino data constrains
solar density fluctuations to be less than $\beta = 0.05$ at the 70 \%
confidence level. The best fit
to the combined solar neutrino and KamLAND data is given by $\beta =
0$ (exact SSM).

In the effective field theory approach to nuclear interactions,  
nonlocal interactions at
short distances are represented by effective local interactions
in a derivative expansion. Since the effect of a given operator on
low-energy physics is inversely proportional to its dimension,
an effective theory valid at low energies can be
written down by retaining operators up to a given dimension. It turns out that 
one needs to introduce a single coefficient,
commonly called $ L_{1A} $, to parameterize the unknown
isovector axial two-body current which dominates the
uncertainties of all neutrino-deuteron interactions \cite{Butler:1999sv}. 
Chen, Heeger, and Robertson, using the Sudbury Neutrino Observatory (SNO)
and SuperKamiokande (SK) charged-current, neutral current, and
elastic scattering rate data, 
found \cite{Chen:2002pv} $L_{1A}  = 4.0 \pm 6.3 \: \mathrm{fm}^3$. 
In order to obtain this result they wrote the observed
rate in terms of an averaged effective cross section and a
suitably defined response function. One can explore
the phenomenology associated with the variation of $L_{1A} $. For example 
the variation of the neutrino parameter space, which fits the SNO data, as 
$L_{1A} $ changes 
was calculated in \cite{Balantekin:2003ep} and is shown 
in Figure 3. In Ref. \cite{Balantekin:2003ep} the 
most conservative fit value with fewest assumptions is found to be 
$ L_{1A} = 4.5 ^{+18}_{-12} \: \mathrm{fm}^3$. 
It was also shown that the contribution of the uncertainty of $ L_{1A} $ to
the analysis and interpretation of the solar neutrino data measured
at the Sudbury Neutrino Observatory is significantly less than the
uncertainty coming from the lack of having a better knowledge
of $ \theta_{13} $, the mixing angle between first and third generations. 

In conclusion we would like to reiterate that the utility of the solar neutrino 
and related reactor and long-baseline neutrino experiments goes well beyond 
that of exploring neutrino parameter space. In this short note we briefly discussed 
only two of such applications out of a much longer list.   

This work was supported in part by the U.S. National Science
Foundation Grant No.\ PHY-0244384 and in part by
the University of Wisconsin Research Committee with funds granted by
the Wisconsin Alumni Research Foundation.

\end{document}